\documentclass[12pt]{article}
\pdfoutput=1

\usepackage{amsmath,hyperref}
\usepackage{fullpage}

\newtheorem{theorem}{Theorem}

\begin{document}

\title{Nonexistence of supersymmetry breaking\\
       counterexamples to the Nelson-Seiberg theorem}
\author{Zhenhuan Li\textsuperscript{*}, Zheng Sun\textsuperscript{\dag}\\
        \normalsize\textit{College of Physics, Sichuan University,}\\
        \normalsize\textit{29 Wangjiang Road, Chengdu 610064, P.~R.~China}\\
        \normalsize\textit{E-mail:}
        \textsuperscript{*}\texttt{985898250@qq.com,}
        \textsuperscript{\dag}\texttt{sun\_ctp@scu.edu.cn}
       }
\date{}
\maketitle

\begin{abstract}
Counterexample models to the Nelson-Seiberg theorem have been discovered, and their features have been studied in previous literature.  All currently known counterexamples have generic superpotentials respecting the R-symmetry, and more R-charge $2$ fields than R-charge $0$ fields.  But they give supersymmetric vacua with spontaneous R-symmetry breaking, thus violate both the Nelson-Seiberg theorem and its revisions.  This work proves that the other type of counterexamples do not exist.  When there is no R-symmetry, or there are no more R-charge $2$ fields than R-charge $0$ fields in models with R-symmetries, generic superpotentials always give supersymmetric vacua.  There exists no specific arrangement of R-charges or non-R symmetry representations which makes a counterexample with a supersymmetry breaking vacuum.  This nonexistence theorem contributes to a refined classification of R-symmetric Wess-Zumino models.
\end{abstract}

\section{Introduction}

The Nelson-Seiberg theorem~\cite{Nelson:1993nf} and its revisions~\cite{Kang:2012fn, Li:2020wdk} relate R-symmetries to F-term supersymmetry (SUSY) breaking~\cite{Intriligator:2007cp} in Wess-Zumino models~\cite{Wess:1973kz, Wess:1974jb} or O'Raifeartaigh models~\cite{ORaifeartaigh:1975nky}.  In addition, R-symmetric SUSY vacua can also be obtained with certain R-charge assignments~\cite{Sun:2011fq, Brister:2021xxxx}.  These relations are applied in the study of both SUSY phenomenology beyond the Standard Model (SM)~\cite{Nilles:1983ge, Martin:1997ns, Baer:2006rs, Terning:2006bq, Dine:2007zp} and compactification of string theory~\cite{Grana:2005jc, Douglas:2006es, Blumenhagen:2006ci, Ibanez:2012zz, Blumenhagen:2013fgp}.  However, the recently discovered counterexample models~\cite{Sun:2019bnd, Amariti:2020lvx} have vacuum solutions contradicting with the claims from the Nelson-Seiberg theorem and its revisions.  These counterexamples have generic coefficients respecting symmetries and do not suffer from fine-tuning, thus are not neglectable in phenomenology studies.  Their features are summarized into a sufficient condition~\cite{Sun:2021svm}, which produces a class of models covering all known counterexamples in the previous literature.

All counterexamples found so far have more R-charge $2$ fields than R-charge $0$ fields.  Their generic superpotentials lead to SUSY vacua with spontaneous R-symmetry breaking.  Thus both the Nelson-Seiberg theorem and its revisions are violated.  It is natural to ask whether the other type of counterexamples exist or not, and this work proves their nonexistence.  When there is no R-symmetry, or there are no more R-charge $2$ fields than R-charge $0$ fields, superpotentials with generic coefficients respecting symmetries always give SUSY vacua.  Additional symmetries which are not R-symmetries can also be considered, and the nonexistence result is unchanged.  Combined with previous theorems relating R-symmetries to SUSY breaking, this nonexistence theorem enables a refined classification of R-symmetric Wess-Zumino models.

The rest part of this work is as following.  Section 2 reviews the Nelson-Seiberg theorem, its revisions and the sufficient condition for counterexamples.  Section 3 gives the proof for the nonexistence of SUSY breaking counterexamples.  Section 4 makes conclusion and remarks.

\section{The Nelson-Seiberg theorem and counterexamples}

A Wess-Zumino model has a holomorphic superpotential $W(z_i)$ of chiral superfields or their scalar components $z_i$'s.  SUSY vacua correspond to solutions to the F-term equations
\begin{equation}
F_i^* = \partial_i W
      = \frac{\partial W}{\partial z_i}
      = 0. \label{eq:2-1}
\end{equation}
They are also minima of the scalar potential
\begin{equation}
V = K^{\bar{i} j} (\partial_i W)^* \partial_j W,
\end{equation}
which involves the positive-definite K\"ahler metric $K^{\bar{i} j}$.  SUSY breaking vacua correspond to minima of $V$ which are not solutions to \eqref{eq:2-1}.  If runaway directions~\cite{Ferretti:2007ec, Ferretti:2007rq, Azeyanagi:2012pc, Sun:2018hnk} are also viewed as prototypes of SUSY breaking vacua before introducing corrections at large field values, the nonexistence of a solution to~\eqref{eq:2-1} can be used as a criteria for SUSY breaking.  Conditions for SUSY breaking are found in terms of R-symmetries under which $W$ has R-charge $2$, summarized as the following theorems~\cite{Nelson:1993nf, Kang:2012fn, Li:2020wdk}:

\begin{theorem} \label{thm:01}
(The Nelson-Seiberg theorem)
In a Wess-Zumino model with a generic superpotential, an R-symmetry is a necessary condition, and a spontaneously broken R-symmetry is a sufficient condition for SUSY breaking at the true vacuum.
\end{theorem}

\begin{theorem} \label{thm:02}
(The Nelson-Seiberg theorem revised and generalized)
In a Wess-Zumino model with a generic superpotential, SUSY is spontaneously broken at the true vacuum if and only if the superpotential has an R-symmetry, and one of the following conditions is satisfied:
\begin{itemize}
    \item The superpotential is smooth at the origin of the field space, and the number of R-charge $2$ fields is greater than the number of R-charge $0$ fields for any consistent R-charge assignment.
    \item The superpotential is singular at the origin of the field space.
\end{itemize}
\end{theorem}

In addition, the existence and properties of R-symmetric SUSY vacua are described by the following theorem~\cite{Sun:2011fq}:

\begin{theorem} \label{thm:03}
(A sufficient condition for R-symmetric SUSY vacua)
In a Wess-Zumino model with a generic R-symmetric superpotential, a sufficient condition for a SUSY vacuum is that the number of R-charge $2$ fields is less than or equal to the number of R-charge $0$ fields for a consistent R-charge assignment.  The obtained SUSY vacuum has its degeneracy dimension equal to the difference between the number of R-charge $2$ fields and the number of R-charge $0$ fields.  The R-symmetry is preserved everywhere on the vacuum.  And the superpotential has the expectation value $W = 0$.
\end{theorem}

While these relations have found successful applications in SUSY phenomenology beyond SM and compactification of string theory, counterexamples to the Nelson-Seiberg theorem and its revisions are constructed in literature, with generic superpotential coefficients and non-generic R-charges~\cite{Sun:2019bnd, Amariti:2020lvx}.  Their features are summarized into the following sufficient condition~\cite{Sun:2021svm}:

\begin{theorem} \label{thm:04}
(A sufficient condition for counterexamples)
In a Wess-Zumino model with a generic R-symmetric superpotential, we classify fields according to their R-charges into the following types:
\begin{itemize}
    \item $X_i$ fields for $i = 1, \dotsc, N_X$, with $R(X_i) = 2$;
    \item $Y_i$ fields for $i = 1, \dotsc, N_Y$, with $R(Y_i) = 0$;
    \item $P_{(r) i}$ and $Q_{(- r) j}$ fields for $i = 1, \dotsc, N_{P(r)}$ and $j = 1, \dotsc, N_{Q(- r)}$, with $R(P_{(r) i}) = - R(Q_{(- r) j}) = r$  for a certain positive value of $r$, and both $P$'s and $Q$'s only appear linearly in the superpotential and not in any quadratic terms;
    \item $A_i$ fields for $i = 1, \dotsc, N_A$, with $R(A_i) = r_i \notin \{ 2, 0 \}$, and $A$'s do not satisfy the condition for being classified as $P_{(r)}$ and $Q_{(- r)}$ fields.
\end{itemize}
Then $N_Y < N_X \le N_Y + N_{P Q}$ is a sufficient condition for the model to be a counterexample to both the Nelson-Seiberg theorem and its revisions, where $N_{P Q} = \sum_r \left ( N_{P(r)} + N_{Q(- r)} - 1 \right )$ is the number of independent $P$-$Q$ pairs.  Such a model gives a SUSY vacuum with degeneracy dimension equal to $N_Y + N_{P + Q} - N_X$, where $N_{P + Q} = \sum_r \left ( N_{P(r)} + N_{Q(- r)} \right )$ is the total number of $P$'s and $Q$'s.  The R-symmetry is spontaneously broken by the non-zero VEV's of $P$-$Q$ pairs everywhere on the degenerated vacuum.  And the superpotential has the expectation value $W = 0$.
\end{theorem}

All counterexamples found so far are covered by this sufficient condition.  The simplest one~\cite{Sun:2019bnd} has four fields with the R-charge assignment
\begin{equation}
(R(X), R(P), R(Q), R(A)) = (2, 6, -6, -2).
\end{equation}
The superpotential with all renormalizable R-charge $2$ terms is
\begin{equation}
W = a X + d X P Q + \xi X^2 A + \sigma P A^2,
\end{equation}
where $a$, $d$, $\xi$ and $\sigma$ are generic complex-valued coefficients.  The F-term equations have a SUSY solution at
\begin{equation}
X = A
  = 0, \quad
P Q = - a / d
\end{equation}
with $W = 0$.  This model violates Theorem \ref{thm:01} and \ref{thm:02}, with its non-generic R-charge assignment satisfying Theorem \ref{thm:04}.  Unlike the R-symmetric SUSY vacua in Theorem \ref{thm:03}, here the obtained SUSY vacuum breaks the R-symmetry for any nonzero values of $a$ and $d$.  Thus the counterexamples covered by Theorem \ref{thm:04} form a new class of R-symmetric Wess-Zumino models beside the previously known classification by Theorem \ref{thm:02} and \ref{thm:03}.  It is natural to ask whether there are other types of counterexamples.  To partially answer this question, nonexistence of counterexamples with SUSY breaking vacua is proved in the next section.

\section{Nonexistence of SUSY breaking counterexamples}

Counterexamples in the context of this work are generic models violating Theorem \ref{thm:01} and \ref{thm:02}.  The term ``generic'' means that all renormalizable terms respecting symmetries must be present in the superpotential, with their coefficients taking generic complex values.  The obtained solutions, either SUSY or SUSY breaking, must be robust against small perturbations of coefficients.  The counterexamples with SUSY vacua covered by Theorem \ref{thm:04} fit this context of genericness, and we have abandoned the notion of "non-generic R-charges" in the following discussion.  To be a SUSY breaking counterexample to Theorem \ref{thm:01} and \ref{thm:02}, either the model has no R-symmetry, or the number of R-charge $2$ fields is less than or equal to the number of R-charge $0$ fields for a consistent R-charge assignment in the model with an R-symmetry.  Theorem \ref{thm:01} and \ref{thm:02} in these cases predict the existence of SUSY vacua, which are absent in the possible SUSY breaking counterexamples.  These cases will be checked in the following proof.
\begin{itemize}
    \item If the model has no R-symmetry, the superpotential takes the most general form without any symmetry restriction.  It includes all terms up to cubic with generic coefficients:
         \begin{equation}
         W = w + a_i z_i + b_{i j} z_i z_j + c_{i j k} z_i z_j z_k.
         \end{equation}
         The F-term equations
         \begin{equation}
         \partial_i W = a_i + 2 b_{i j} z_j + 3 c_{i j k} z_j z_k
         \end{equation}
         form a balanced system with $N$ variables and $N$ quadratic polynomial equations.  There are $\frac{1}{6} (N + 1)(N + 2)(N + 3) - 1$ coefficients taking generic complex values.  According to Hilbert's Nullstellensatz, nonexistence of a solution to these equations means that values of coefficients must be tuned to a subspace of the whole space of coefficients.  Models with such fine-tuned coefficients are non-generic, and fall out of the scope of this work.  So generic models in this case always yield SUSY vacua.
    \item If the model has an R-symmetry, and the number of R-charge $2$ fields is less than or equal to the number of R-charge $0$ fields for a consistent R-charge assignment, the most generic form of the superpotential is~\cite{Sun:2011fq}
         \begin{equation}
         \begin{aligned}
         W &= W_0 + W_1,\\
         W_0 &= X_i (a_i + b_{i j} Y_j + c_{i j k} Y_j Y_k),\\
         W_1 &= \underbrace{\xi_{i j} X_i^2 A_j}_{r_j = - 2}
                + \underbrace{\rho_{i j k} X_i A_j A_k}_{r_j + r_k = 0}
                + \underbrace{(\mu_{i j} + \nu_{i j k} Y_k) A_i A_j}_{r_i + r_j = 2}
                + \underbrace{\lambda_{i j k} A_i A_j A_k}_{r_i + r_j + r_k = 2},
         \end{aligned}
         \end{equation}
         where fields are classified to three types:  $X_i$ fields with R-charge $2$, $Y_i$ fields with R-charge $0$, and $A_i$ fields with R-charges $R(A_i) = r_i \notin \{ 2, 0 \}$.  If we look for vacua in the condition
         \begin{equation}
         X_i = A_j
             = 0, \label{eq:3-1}
         \end{equation}
         the following first derivatives vanish at such vacua:
         \begin{equation}
         \partial_{X_i} W_1 = \partial_{Y_j} W_1
                            = \partial_{A_k} W_1
                            = 0, \quad
         \partial_{Y_i} W_0 = \partial_{A_j} W_0
                            = 0.
         \end{equation}
         The F-term equations are then reduced to
         \begin{equation}
         \partial_{X_i} W_0 = a_i + b_{i j} Y_j + c_{i j k} Y_j Y_k
                            = 0. \label{eq:3-2}
         \end{equation}
         They form a balanced or under-determined system with $N_Y$ variables and $N_X \le N_Y$ quadratic polynomial equations, where $N_X$ and $N_Y$ are the numbers of $X_i$ and $Y_i$ fields.  There are $\frac{1}{2} N_X (N_Y + 1)(N_Y + 2)$ coefficients taking generic complex values.  Similarly to the previous case, nonexistence of a solution to these equations means that values of coefficients must be tuned to a subspace of the whole space of coefficients.  Models with such fine-tuned coefficients are non-generic, and fall out of the scope of this work.  Generic models in this case always yield solutions to \eqref{eq:3-2}.  So SUSY vacua satisfying \eqref{eq:3-1} always exist, and there is no need to look for vacua not satisfying \eqref{eq:3-1}.
\item If the model has some other symmetries which are not R-symmetries, every field falls in an irreducible representation of every symmetry.  We classify field into two types:  $z_i$ fields which are in the trivial representations of all non-R symmetries, and $z'_i$ fields for others.  The superpotential is written as
      \begin{equation}
      W_\text{total} = W_\text{trivial}(z_i) + W'(z_i, z'_i),
      \end{equation}
      where $W_\text{trivial}$ includes all terms not depending on $z'_i$ fields, and every term in $W'$ depends on some $z'_i$.  To keep the superpotential invariant under all non-R symmetries, $z'_i$ fields can only appear in quadratic or higher order terms of $W'$.  If we look for vacua satisfying
      \begin{equation}
      z'_i = 0, \label{eq:3-3}
      \end{equation}
      which leads to
      \begin{equation}
      \partial_{z_i} W' = \partial_{z'_i} W'
                        = 0,
      \end{equation}
      the F-term equations are then reduced to
      \begin{equation}
      \partial_{z_i} W_\text{trivial} = 0. \label{eq:3-4}
      \end{equation}
      Note that any combination of $z_i$ fields is invariant under non-R symmetries.  Thus non-R symmetries put no restriction on the form of $W_\text{trivial}$.  The previous procedure continues with $W$ replaced by $W_\text{trivial}$.  In both cases we obtain solutions to \eqref{eq:3-4} in generic models, so SUSY vacua satisfying \eqref{eq:3-3} always exist.
\end{itemize}

In summary, we have proved that SUSY vacua predicted by Theorem \ref{thm:01} and \ref{thm:02} always exist in generic models, so SUSY breaking counterexamples do not exist:

\begin{theorem} \label{thm:05}
(Nonexistence of SUSY breaking counterexamples)
In a Wess-Zumino model with generic superpotential coefficients respecting symmetries, the following two cases always yield SUSY vacua:
\begin{itemize}
\item The model has no R-symmetry.
\item The model has an R-symmetry, and the number of R-charge $2$ fields is less than or equal to the number of R-charge $0$ fields for a consistent R-charge assignment.
\end{itemize}
The model can also have some non-R symmetries which do not change the result.  There is no special arrangement of R-charges or non-R symmetry representations for a counterexample with SUSY breaking true vacua to exist.
\end{theorem}

\section{Conclusion}

In this work, we have proved the nonexistence of SUSY breaking counterexamples to the Nelson-Seiberg theorem and its revisions, expressed as Theorem \ref{thm:05}.  Thus all counterexamples with generic superpotential coefficients are of the SUSY type, either or not covered by Theorem \ref{thm:04}.  If we use the field counting criteria from Theorem \ref{thm:02} to classify out SUSY vacua with $W = 0$, which contribute to the low energy SUSY branch of the string landscape~\cite{Dine:2004is, Dine:2005yq, Dine:2005gz}, the nonexistence of SUSY breaking counterexamples means that all the SUSY counterexamples are false negatives, which affect the recall but retain the precision of the classifier.  Combined with Theorem \ref{thm:04} describing all currently known SUSY counterexamples, our nonexistence theorem contributes to a refined classification of R-symmetric Wess-Zumino models.

The technique in the proof of this work is also applicable in the proof of Theorem \ref{thm:04}.  The field classification in Theorem \ref{thm:04} leads to the generic superpotential~\cite{Sun:2021svm}
\begin{equation}
\begin{aligned}
W &= W_0 + W_1,\\
W_0 &= X_i (a_i + b_{i j} Y_j + c_{i j k} Y_j Y_k
            + d_{(r) i j k} P_{(r) j} Q_{(- r) k}),\\
W_1 &= \underbrace{\xi_{i j} X_i^2 A_j}_{r_j = - 2}
       + \underbrace{\rho_{i j k} X_i A_j A_k}_{r_j + r_k = 0}
       + \underbrace{\sigma_{(r) i j k} P_{(r) i} A_j A_k}_{r_j + r_k = 2 - r}
       + \underbrace{\tau_{(r) i j k} Q_{(- r) i} A_j A_k}_{r_j + r_k = 2 + r}\\
    &\quad
       + \underbrace{(\mu_{i j} + \nu_{i j k} Y_k) A_i A_j}_{r_i + r_j = 2}
       + \underbrace{\lambda_{i j k} A_i A_j A_k}_{r_i + r_j + r_k = 2}.
\end{aligned}
\end{equation}
SUSY vacua satisfying
\begin{equation}
X_i = A_j
    = 0
\end{equation}
can be found by solving the reduced F-term equations
\begin{equation}
a_i + b_{i j} Y_j + c_{i j k} Y_j Y_k + d_{(r) i j k} P_{(r) j} Q_{(- r) k} = 0. \label{eq:4-1}
\end{equation}
These equations form a balanced or under-determined system with effectively $N_Y + N_{P Q}$ independent variables and $N_X \le N_Y + N_{P Q}$ quadratic polynomial equations.  There are $\frac{1}{2} N_X ((N_Y + 1)(N_Y + 2) + N'_{P Q})$ coefficients taking generic complex values, where $N'_{P Q} = \sum_r N_{P(r)} N_{Q(- r)}$ is the total number of $P$-$Q$ pairs.  Similarly to the previous proof, one can always find solutions to \eqref{eq:4-1} with generic coefficients.  So generic models satisfying Theorem \ref{thm:04} always give SUSY vacua without exceptions.

\section*{Acknowledgement}

We thank Yan He for helpful discussions.  This work is supported by the National Natural Science Foundation of China under grant 11305110.

\end{document}